\documentclass[preprint,preprintnumbers]{revtex4}

\usepackage{graphicx}

\begin{document}
\title { A microscopic model  for detecting  the  surface states  photoelectrons    in Topological Insulators  }

\author{ D.Schmeltzer}

\affiliation{Physics Department, City College of the City University of New York \\
New York, New York 10031}

\begin{abstract}

We present a model for  the photoelectrons emitted from the surface of a Topological insulator  induced by a  polarized  laser source. The model is based on the tunneling of the surface electrons  into the vacuum in the presence of a photon field. Using the Hamiltonian  which describes the  coupling of the photons  to the surface electrons we compute   the intensity and polarization of the photoelectrons. 

\end{abstract}

\maketitle
Recently  the use of an innovative  spectrometer with a high laser-based light source  have shown that the spin polarization of the  photoelectrons  emitted from the surface of the  $ Bi_{2}Se_{3}$  Topological Insulator ($T.I.$) \cite{Volkov,Zhang,Kane,David}  can be manipulated  through the  laser light  polarization \cite{Nature}. One finds that the photoelectrons  polarization  is completely  different  from the  initial states  and is controlled by the  photon polarization. A number of theories have been proposed \cite{Park,Wang,Moore}.
A correct  theory for this effect is important since the photoemission experiments   are used to investigate the nature of the surface states.
The theory given by  \cite{Park} suggest that the photoelectron current  with the spin part   in the $  \hat{t}$  direction  is given  by $ \vec{S}\cdot \hat{t}$  where $ \vec{S}$ is the spin polarization of the photoelectrons.  The state recorded by the photodetector   at position  $\vec{R}_{d}$ is given by $|\hat{t} ,\vec{R}_{d}>$.   The  photocurrent  induced  from the intial  T.I. state $ |i>$ depends on the  photoexcited states $ |f>$ of the Topological Surface States. ($T.S.S.$) is given  in ref.  \cite{Park} :
$I_{\hat{t}}\propto |\sum_{E_{f}=E_{i}+\hbar \Omega}<\hat{t},,\vec{R}_{d}|f><f|H^{int}|i>|^2$ where $ H^{int}$ is the matrix element between the initial surface electron and final state and $ \langle\hat{t},\vec{R}_{d}|f\rangle$ is the projection of the final state $|f\rangle$  measured by the detector. 
This phenomenological description is based   the Fermi  Golden rule  for  the photoemission cross section.  
Recently it has been suggested  \cite{Lin} that interactions and hybridization between the bands in a T.I. might be responsible for the modification of the  photoemission spectrum.

The purpose of this  paper is to present a microscopic model which   explains the photoemission  experimental results. We present a model  where the surface $ T.I.$ conduction band are coupled to the  normal electrons of the vacuum states.  Such a model emerges when we consider the bulk $T.I.$ electrons which give rise to the surface states , the surface states have a non-zero amplitude to tunnel into the vacuum \cite{Fan,David} .
 The  basic  $T.S.S.$ Hamiltonian in the presence of the vector potential  $ \vec{A} $ is given by  the Weyl model \cite{D.S.}:
\begin{eqnarray}
&&
H=\hbar v  \int\,d^2 r \Psi^{\dagger}(\vec{r}) \left[\sigma^{1}\left(-i\partial_{2}-\frac{e}{\hbar}A_{2}(\vec{r},t)\right)-\sigma^{2}\left(-i\partial_{1}-\frac{e}{\hbar}A_{1}(\vec{r},t)\right)\right]\Psi(\vec{r}) , \nonumber\\&& 
\end{eqnarray}
The spinor operator $\Psi(\vec{r})$  is decomposed into the eigenvalues  of the unperturbed  Weyl  Hamiltonian  
\begin{equation}
\Psi(\vec{r})=\sqrt{v}\sum_{\vec{k}}e^{i\vec{k}\cdot \vec{r}}[C(\vec{k})u^{(+)}(\vec{k})+ B^{\dagger}(-\vec{k})u^{(-)}(-\vec{k})]\equiv \sqrt{v} \sum_{\vec{k}}e^{i\vec{k}\cdot \vec{r}}\Psi(\vec{k}) . 
\label{gr}
\end{equation}
where   $|u^{(+)}(\vec{k})\rangle$  is  the eigenvector  for particles (positive  energies)  and $|u^{(-)}(\vec{k})\rangle$ is the spinor for  antiparticles  (negative energies) : 
\begin{eqnarray}
&&|u^{(-)}(\vec{k})\rangle=\frac{1}{\sqrt{2}}\Big[|\vec{k}\rangle\otimes[ie^{-i\chi(\vec{k})},1]^{T}\Big] . 
\label{state} ,\nonumber\\&&
|u^{(+)}(\vec{k})\rangle=\frac{1}{\sqrt{2}}\Big[|\vec{k}\rangle\otimes[1,-ie^{i\chi(\vec{k})}]^{T}\Big], \hspace{0.2 in} . 
\end{eqnarray}
The eigenvectors  are given in terms of the multivalued phase  $\chi(\vec{k})$, $\tan(\chi(\vec{k}))=\frac{k_{2}}{k_{1}}$.
We choose for   the particle operators $C(\vec{k})$, $C^{\dagger}(\vec{k})$   and for  the anti-particles the  operators  $B(\vec{k})$, $B^{\dagger}(\vec{k})$.
 In order to construct the explicit model we will take  the   chemical potential $ \mu=0.3$ ev.
 We can consider only the particle band, electrons excitations above the Dirac Cone.The momentum $\vec{k}$ is the momentum parallel to the surface located at $z=0$. For simplicity we have ignored the hexagonal warping effect  on the energy spectrum of the quasi-particles \cite{Liang} and consider only the effect  on the spinors  \cite{D.S.}.

Next we present the three parts   of the  model : A- the one band  Hamiltonian and the  coupling to the  photon  field   ; B-The coupling of the surface electrons to the vacuum electrons ;  C-The spin detection Hamiltonian:

\textbf{A- The one band  Hamiltonian and the  coupling to the external photon field   is given by}:
\begin{equation}
H^{(c)}=\sum_{\vec{k}}(\hbar v|\vec{k}|-\mu)C^{\dagger}(\vec{k},z=0)C(\vec{k},z=0)
\label{conduction}
\end{equation}
The Hamiltonian $ H^{(c)}$ with the chemical potential $\mu$ describe the surface conduction electrons localized on the surface $z=0$.  In agreement with ref. \cite{D.S.} we only consider the conduction electrons and ignore the antiparticles excitations.
The coupling to the photon field is given according to ref.\cite{D.S.} and depends explicitly on the matrix elements  $\langle u^{(+)}(\vec{k})|\sigma^1 |u^{(+)}(\vec{p})\rangle$ and $ \langle u^{(+)}(\vec{k})|-\sigma^2 |u^{(+)}(\vec{p})\rangle$.
\begin{eqnarray}
&&H^{ext}(t)=\int\ d^2 r[ J_{1}(\vec{r},t)A_{1}(\vec{r},t) + J_{2}(\vec{r},t)A_{2}(\vec{r},t)]=(-ev)\int\frac{d^2k}{(2\pi)^2}\int\frac{d^2p}{(2\pi)^2}\nonumber\\&&C^{\dagger}(\vec{k},t)C(\vec{p},t)[ \langle u^{(+)}(\vec{k})|-\sigma^2 |u^{(+)}(\vec{p})\rangle A_{1}(-(\vec{k}-\vec{p}),t)+\langle u^{(+)}(\vec{k})|\sigma^1 |u^{(+)}(\vec{p})\rangle A_{2}(-(\vec{k}-\vec{p}),t)]\nonumber\\&&=(-ev)\int\frac{d^2k}{(2\pi)^2}\int\frac{d^2p}{(2\pi)^2}\hat{C}^{\dagger}(\vec{k},t)\hat{C}(\vec{p},t)\Big[W_{x}(\vec{k},\vec{p})A_{x}   (-(\vec{k}-\vec{p}),t)+W_{y}(\vec{k},\vec{p})A_{y}(-(\vec{k}-\vec{p}),t)\Big]\nonumber\\&&
W_{x}(\vec{k},\vec{p})= \cos[\frac{1}{2}(\chi(\vec{k})+\chi(\vec{p}))];  W_{y}(\vec{k},\vec{p})=-\sin[\frac{1}{2}(\chi(\vec{k})+\chi(\vec{p}))]\nonumber\\&&
\end{eqnarray}
The photon field $ \vec{A}(\vec{P}\equiv\vec{p}-\vec{k},p_{z}) $  characterized by the coherent state $ |\Omega>$. The   direction of the incoming  photon  with respect the surface located at $z=0$  is given by the vector $\frac{ \vec{P}}{|\vec{P}|}$. The    two transversal polarization are   given by the vectors $ \vec{e}_{s=1}(\vec{P})$ and  $\vec{e}_{s=2}(\vec{P})$  which are perpendicular to the photon  propagation.
\begin{eqnarray}
&& \vec{A}(\vec{P},t)=\sum_{s=1,2}[\vec{e_{s}}(\vec{P})a_{s}(\vec{P})e^{-i\Omega t}+
\vec{e_{s}}(\vec{P})a^{\dagger}_{s}(\vec{P)}e^{i\Omega t}]\nonumber\\&&
\frac{ \vec{P}}{|\vec{P}|}= (sin[\theta]cos[\phi],sin[\theta]sin[\phi],cos[\theta])\nonumber\\&&  
\vec{e}_{s=1}(\vec{P})=(cos[\theta]cos[\phi],cos[\theta]sin[\phi],-sin[\theta])\nonumber\\&&
\vec{e}_{s=2}(\vec{P})=(-sin[\phi],cos[\phi],0)\nonumber\\&&
\end{eqnarray}
The coherent state condition for the photon field with the frequency $\Omega$ is given by:$ a_{s}(\vec{P}) |\Omega>=\hat{a}_{s}(\vec{P})|\Omega\rangle$. For the remaining part the photon momentum will be neglected in comparison with the electrons quasi-momentum.  

\textbf{B-The hybridization   between the surface electrons   and emitted  vacuum electrons.}

We will assume that the electrons in the vacuum  are described by the creation and annihilation operators $ b^{\dagger}_{\alpha}(\vec{k},k_{z})$ and  $b_{\alpha}(\vec{k},k_{z})$  with    the single particles  energy, 
$E_{f}(\vec{k},k_{z})=\frac{\hbar^2 K^2}{2 m}+V_{0}$ , where  $V_{0}$ is the binding- work function for  the $T.S.S.$ surface. The tunneling Hamiltonian between the surface at $z=0$ and the vacuum is given by:
\begin{eqnarray}
&&H^{z=0}_{t}=\int\frac{d^2k }{(2\pi)^2}\int\frac{dk_{z_{1}}}{2\pi}\int\frac{dk_{z_{1}}}{2\pi}[t(k_{z_{1}},k_{z_{2}})C_{\alpha}^{\dagger}(\vec{k},k_{z_{1}})b_{\alpha}(\vec{k},k_{z_{2}})+h.c.]\nonumber\\&&
C_{\alpha}(\vec{k},k_{z_{1}})\equiv
C(\vec{k},k_{z_{1}})u_{\alpha}^{(+)}(\vec{k});  u_{\alpha}^{(+)}(\vec{k})\equiv \langle\alpha|u^{(+)}(\vec{k})\rangle \nonumber\\&&
H_{0}=\sum_{\vec{k},k_{z}}(\frac{\hbar^2 (k^2+k^2_{z})}{2 m}+V_{0}-\mu)b_{\alpha}^{\dagger}(\vec{k},k_{z})b_{\alpha}(\vec{k},k_{z})\nonumber\\&&
\end{eqnarray}
 $t(k_{z_{1}},k_{z_{2}})$ is the tunneling matrix element between the surface and the vacuum. The   momentum is not conserved in the $z$ direction.
The diagonal Hamiltonian is  given in eq.$(7)$.:
\begin{eqnarray}
&& C_{\alpha}(\vec{K})=\cos[\eta]\psi_{\alpha}(\vec{K})+   \sin[\eta]B_{\alpha} (\vec{K});
b_{\alpha}(\vec{K})=\cos[\eta]B_{\alpha} (\vec{K})   - \sin[\eta] \psi_{\alpha}(\vec{K})\nonumber\\&&
\tan[\eta]=-\frac{\sqrt{2}t(k_{z},k_{z})}{E_{f}(\vec{k},k_{z})-\hbar |\vec{k}|}\nonumber\\&&
H=\int\frac{d^2k }{(2\pi)^2}\int\frac{dk_{z}}{2\pi}\Big[ (E_{-}(\vec{k},k_{z})-\mu) \psi^{\dagger}_{\alpha}      (\vec{k},k_{z})\psi_{\alpha}(\vec{k},k_{z}) +  (E_{+}(\vec{k},k_{z})-\mu) B^{\dagger}_{\alpha}(\vec{k},k_{z})B_{\alpha}(\vec{k},k_{z})\Big]\nonumber\\&&
E_{\pm}(\vec{k},k_{z})=\frac{1}{2}[\hbar v|\vec{k}|+\frac{\hbar^2 (k^2+k^2_{z})}{2 m}+V_{0}]\pm \Delta(\vec{k},k_{z})\sqrt{[1+(\frac{t}{\sqrt{2}\Delta(\vec{k},k_{z})})^2]}\nonumber\\&& \Delta(\vec{k},k_{z})\equiv \frac{1}{2} [\frac{\hbar^2 (k^2+k^2_{z})}{2 m}+V_{0}-\hbar v|\vec{k}|]\nonumber\\&&
\end{eqnarray}

\textbf{C-The spin detection Hamiltonian.}

The detector is characterized by the vector $\hat{t}$ which in the frame of the $T.S.S.$ is given by the three  vector components:

$\vec{\hat{t}}=  (\hat{t}_{x},\hat{t}_{y},\hat{t}_{z})=(\sin[\theta_{d}]cos[\phi_{d}],\sin[\theta_{d}]sin[\phi_{d}],\cos[\theta_{d}])$.
 
 The momentum  parallel to the surface of the  $T.I.$   is conserved therefore the angle $\phi_{d}$ determines the  phase $\chi(\vec{k})$ of the T.S.S. which is measured, $\cos[\phi_{d}]=\cos[\chi(\vec{k})]$.
The detector Hamiltonian  measures the polarization energy  as function of the local  magnetization    $\vec{ M} $.  

\textbf{Computation of the Intensity of the polarized  Photoelectrons}

The number of polarized photoelectrons  measured in the $ \vec{\hat{t}}$ direction  is given by: 
$I_{\hat{t}}=\langle g|[\hat{t}_{r} b_{\alpha}^{\dagger}(\vec{Q})[\sigma^{(r)}]_{\alpha,\beta}b_{\beta}(\vec{Q})]|g\rangle d^3Q)$,
 $ |g\rangle $ is  the perturbed  ground state  given in terms of the matrix $S=e^{-i\int_{-\infty}^{\infty}H^{ext}(t') dt'}$   \cite{Doniach} and the vacuum state $ |O\rangle\equiv|E_{+},E_{-}\rangle $( obtained  from eq.$(8)$).   We  find  :
\begin{eqnarray}
&& I_{\hat{t}}(\vec{Q},\omega)=\int_{-\infty}^{\infty}\,dt e^{i\omega t}\langle g|[\hat{t}_{r} e^{i\omega t} b^{\dagger}_{\alpha}(\vec{Q},t)[\sigma^{(r)}]_{\alpha,\beta}b_{\beta}(\vec{Q},t)]|g\rangle d^3Q=I^{(1)}_{\hat{t}}(\vec{Q},\omega)+I^{(2)}_{\hat{t}}(\vec{Q},\omega)...\nonumber\\&&
 I^{(1)}_{\hat{t}}(\vec{Q},\omega)= (-i)  \int_{-\infty}^{\infty}\,dt e^{i\omega t}
\int_{-\infty}^{\infty}\,dt_{1}\langle0|T\Big[\hat{t}_{r} b^{\dagger}_{\alpha}(\vec{Q},t)[\sigma^{(r)}]_{\alpha,\beta}b_{\beta}(\vec{Q},t)]H^{ext}(t_{1})\Big]|0\rangle d^3Q\nonumber\\&&
I^{(2)}_{\hat{t}}(\vec{Q},\omega)=\frac{(-i)^{2}}{2}  \int_{-\infty}^{\infty}\,dt e^{i\omega t}
\int_{-\infty}^{\infty}\,dt_{1} \int_{-\infty}^{\infty}\,dt_{2}\langle0|T\Big[\hat{t}_{r} b^{\dagger}_{\alpha}(\vec{Q},t)[\sigma^{(r)}]_{\alpha,\beta}b_{\beta}(\vec{Q},t)H^{ext}(t_{1}) H^{ext}(t_{2})\Big]|0 \rangle\nonumber\\&&
d^3Q\nonumber\\&&
\end{eqnarray}
In order to compute the number of polarized photoelectrons we substitute the operators $ b_{\alpha}(\vec{Q},t)$ and $C(\vec{Q},t)$ in terms of the  new  operators 
$\psi_{\alpha}(\vec{K})$  $B_{\alpha} (\vec{K})$ given in eq.$(8)$.
Using   Wick theorem  \cite{Doniach}  we perform the expectation values for  the intensity operator $ I_{\hat{t}}(\vec{Q},\omega\rightarrow 0)$.
To first order in the laser  field  we find  : $I^{(1)}_{\hat{t}}(\vec{Q},\omega\rightarrow 0)$, this photon amplitude is a function of the coherent time  and time duration of the laser pulse . 
For the coherent state $ |\Omega>$ we have : $a_{s}(\vec{P}) |\Omega>=\hat{a}_{s}(\vec{P})|\Omega>$ where  the amplitude $\hat{a}_{s}(\vec{P}) $  needs  to be averaged over the coherency time or pulse duration  .  As a result this amplitude is  reduced  and $<<\hat{a}_{s}(\vec{P}) >>\approx 0$ with the polarization  $P^{(1)}$,
$P^{(1)}[\chi(\vec{q}),\theta,\theta_{d},\phi,\phi_{d}]=\sum_{r=x,y,z}
   \hat{t}_{r}<\sigma^{r}>W_{r}(\vec{q})e^{r}_{s}$.
 
$I^{(1)}_{\hat{t}}(\vec{Q},\omega\rightarrow 0)\propto <<\hat{a}_{s}(\vec{P})>>
\Big[(\frac{t(k_{z},k_{z}) }{E_{+}(\vec{q},q_{z})  -E_{-}(\vec{q},q_{z})})^2  \frac{f_{F.D.}[E_{-}(\vec{q},q_{z})-\mu] \cdot 2\Gamma}{(E_{+}(\vec{q},q_{z})  -\hbar\Omega -E_{-}(\vec{q},q_{z}))^2+(2\Gamma)^2}\Big]d^3Q
\cdot P^{(1)}[\chi(\vec{k}),\theta,\theta_{d},\phi,\phi_{d}]$.

\textbf{The photoelectron intensity  $I^{(2)}_{\hat{t}}(\vec{Q},\omega\rightarrow 0)$  to second order in the photon field}

The photoelectron intensity , $I^{(2)}_{\hat{t}}(\vec{Q},\omega\rightarrow 0)$  is computed for two different photon polarizations $s=s'=1,2$ as a function of the detector direction  $ \vec{\hat{t}}$. The expectation values are obtained using the averaged Green's function with the lifetime $\Gamma$  \cite{Doniach}  and the  Fermi-Dirac  occupation function.
\begin{eqnarray}
&&I^{(2)}_{\hat{t}}(\vec{Q},\omega\rightarrow 0)\propto
(e v)^2 (\sin[2\eta]\cos[\eta])^2(\hat{a}^{*}_{s} \hat{a}_{s}+\frac{1}{2})
f_{F.D.}(E_{-}(\vec{q},q_{z})-\mu)[1-f_{F.D.}(E_{+}(\vec{q},q_{z})-\mu)]^2 \nonumber\\&&\Big[\frac{1}{(E_{+}(\vec{q},q_{z})-E_{-}(\vec{q},q_{z})-\hbar\Omega)
^{2}+(2\Gamma)^2}[1-\frac{\hbar \Omega}{(E_{+}(\vec{q},q_{z})-E_{-}(\vec{q},q_{z}))(1+    (\frac{2\Gamma}{(E_{+}(\vec{q},q_{z})-E_{-}(\vec{q},q_{z}))})^2}]\nonumber\\&&
+\frac{1}{(E_{+}(\vec{q},q_{z})-E_{-}(\vec{q},q_{z})+\hbar\Omega)
^{2}+(2\Gamma)^2}[1+\frac{\hbar \Omega}{(E_{+}(\vec{q},q_{z})-E_{-}(\vec{q},q_{z}))(1+    (\frac{2\Gamma}{(E_{+}(\vec{q},q_{z})-E_{-}(\vec{q},q_{z}))})^2}]\Big]\nonumber\\&&
P^{(2)}[\chi(\vec{k}),\theta,\theta_{d},\phi,\phi_{d}] d^3Q\nonumber\\&&
\end{eqnarray}
 where  $ P^{(2)}[\chi(\vec{k}),\theta,\theta_{d},\phi,\phi_{d}]\equiv P^{(2)}[\chi(\vec{k}),\vec{e}_{s}(\theta, \phi),\theta_{d},\phi_{d}]$  is the polarization matrix of the photoelectrons.  For the case considered  $\vec{k}=\vec{q}$, the parallel component of the momentum which reaches the detector is equal to the surface momentum $\vec{k}$, therefore we have   $\chi(\vec{k})=\chi(\vec{q})$. 
  The polarization  is given by:
\begin{equation}
  P^{(2)}[\chi(\vec{k}),\vec{e}_{s}(\theta,\phi),\theta_{d},\phi_{d}]=
\Big[\sum_{r=x,y}W_{r}(\vec{k})e^{r}_{s}(\theta,\phi)\Big]^2\sum_{r'=x,y,z} \hat{t}_{r}(\theta_{d},\phi_{d})<(\sigma^{r})^{*}>
\label{p2}
\end{equation}
We will consider first the scalar part of $I^{(2)}_{\hat{t}}(\vec{Q},\omega\rightarrow 0)$  and   second we will study the explicit dependence of the polarization vector $ P^{(2)}[\chi(\vec{k}),\theta,\theta_{d},\phi,\phi_{d}]$.

\textbf{a-The scalar amplitude of the emitted electrons}

Using the formula  $I^{(2)}_{\hat{t}}(\vec{Q},\omega\rightarrow 0)$ we plot the intensity as a function of the surface electron energy $\epsilon(\vec{k})$ for different values of the outgoing momentum which is determined by the angle $ \theta_{d}$.  We  use the chemical potential $ \mu=0.3 ev.$ binding- work function $V_{0}\approx5ev$  and laser energy $\hbar\Omega\approx5ev.$.  The energy of the emitted electrons  can be obtained from the fact that the surface momentum $\vec{k}=q_{||}=|\vec{Q}|\ sin[\theta_{d}]$ and the $ z$ component $q_{z}=|\vec{Q}|\cos[\theta_{d}]$.  If   $\epsilon  $  is the surface electron energy  the free electron energy is given by  $E_{f}(\vec{q},q_{z})= \frac{\hbar^2Q ^2}{2 m}+V_{0 }\equiv\frac{[\frac{ \epsilon}{\sin[\theta_{d}]}]^2}{2 m}+V_{0}$.
The intensity  $I^{(2)}_{\hat{t}}(\vec{Q},\omega\rightarrow 0)$  as a function of surface electrons energy $\epsilon  $   for a fixed angle $\theta_{d}=\frac{\pi}{2}$ in figure $1$  .  In figure $2$  we tune laser frequency and   the binding-work function  to be $5.ev.$. The plots are for  diferent  detector angles $\theta_{d}$ .

\begin{figure}
\begin{center}
\includegraphics[width=6.0 in ]{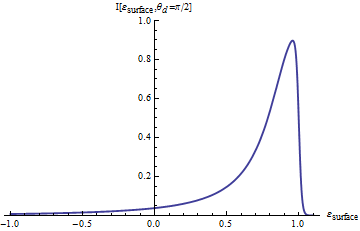}
\end{center}
\caption{The intensity  $I^{(2)}_{\hat{t}}(\vec{Q},\omega\rightarrow 0)$ in arbitrary units   as a function of surface electrons energy $\epsilon  $   for a fixed  angle $\theta_{d}=\frac{\pi}{2}$.      } 
\end{figure}

\begin{figure}
\begin{center}
\includegraphics[width=6.0 in ]{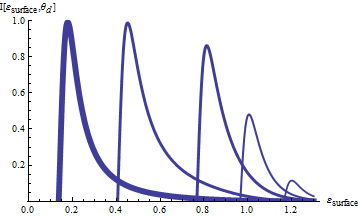}
\end{center}
\caption{The intensity  $I^{(2)}_{\hat{t}}(\vec{Q},\omega\rightarrow 0)$ in arbitrary units   as a function of surface electrons energy $\epsilon  $   for different  detector angle $\theta_{d}$. Laser frequency and the binding-work function have been chosen to be $5ev.$  We plot the intensity for $\theta_{d} =0.87$ (the smallest peak)  and $\theta_{d} =0.27$  corresponds to the peak around  $\epsilon=0.2$.  The values of $\theta_{d}$ considered are   $\theta_{d}=0.87;0.77;0.67; 0.47;0.27$. The surface energies   $\epsilon$  are normalized in units of 0.3 $ ev.$, the bottom of the surface conduction band is at   $\epsilon=0$ .  (1 corresponds to$ 0.3 .$)     } 
\end{figure}

\textbf{b-The polarization of the emitted electrons}

In order to evaluate the polarization of the photoelectrons  we need to use the explicit  form of  the photon polarization vectors $\vec{e}_{s}(\theta,\phi)$ ,$s=1,2$.
In addition we need  the photon  matrix elements $W^{x}(\vec{k})$, $W^{y}(\vec{k})$ and the polarization of the surface  electrons $<\sigma^{x}>$,$<\sigma^{y}>$ .
\begin{equation}
W_{x}(\vec{k})=\cos[\chi(\vec{k})];  W_{y}(\vec{k})=-\sin[\chi(\vec{k})] ;\hspace{0.1 in}
<\sigma^{x}>=\sin[\chi(\vec{k})];  <\sigma^{y}>=-\cos[\chi(\vec{k})] ;  <\sigma^{z}>=0 
\label{eqp}
\end{equation}
Next we consider the case that  the detector is oriented in the  $\vec{\hat{t}}$  direction. 
 The  polarization in the   $\vec{\hat{t}}$  direction is given by :
\begin{equation}
\vec{\hat{t}}\cdot <\vec{\sigma}>=\sin[\theta_{d}]\cos[\chi(\vec{k})+\phi_{d}]=\sin[\theta_{d}]\cos[2\chi(\vec{k})]=\cos[\theta_{d}]\cos[2\phi_{d}]
\label{vectornew}
\end{equation}
Therefore the surface polarization electrons will be given by:
 $ P_{T.S.S.}= \vec{\hat{t}}\cdot< \vec{\sigma}>=\sin[\theta_{d}]\cos[\chi(\vec{k})]$.
For the incoming photons  polarization   $\vec{e}_{s=2}(\theta,\phi)$  we obtain  the photoelectrons polarization $ P^{(2)}[\chi(\vec{k})=\phi_{d},\vec{e}_{s=2}(\theta=0, \phi),\theta_{d},\phi_{d}]$:
\begin{eqnarray}
&& P^{(2)}[\chi(\vec{k})=\phi_{d},\vec{e}_{s=2}(\theta=0, \phi),\theta_{d},\phi_{d}]=\sin[\theta_{d}]\cos[\chi(\vec{k})+\phi_{d}]
\sin^{2}[\phi+\chi(\vec{k})]=\nonumber\\&&\sin[\theta_{d}]\cos[2\chi(\vec{k})]\sin^{2}[\phi+\chi(\vec{k})]\nonumber\\&&
P^{(2)}[\chi(\vec{k})=\phi_{d},\vec{e}_{s=2}(\theta=0, \phi=0),\theta_{d},\phi_{d}]=\sin[\theta_{d}]\cos[2\chi(\vec{k})]\sin^{2}[\chi(\vec{k})]\nonumber\\&&
\end{eqnarray}

The polarization operator $P^{(2)}$ for the photon  polarization $\vec{ e}_{s=1}(\theta,\phi)$ is given by:
\begin{eqnarray}
&& P^{(2)}[\chi(\vec{k})=\phi_{d},\vec{e}_{s=1}(\theta, \phi),\theta_{d},\phi_{d}]= \sin[\theta_{d}]\cos[\chi(\vec{k})+\phi_{d}] \cos^2[\theta]  \cos^2[\chi(\vec{k})+\phi]\nonumber\\&&=\sin[\theta_{d}]  \cos^2[\theta]\cos[2\chi(\vec{k})]\cos^2[\chi(\vec{k})+\phi]
\nonumber\\&&
 P^{(2)}[\chi(\vec{k})=\phi_{d},\vec{e}_{s=1}(\theta, \phi=0),\theta_{d},\phi_{d}]=\sin[\theta_{d}]  \cos^2[\theta]\cos[2\chi(\vec{k})]\cos^2[\chi(\vec{k})]
\nonumber\\&&
\end{eqnarray}
We observe that for th surface polarization electrons  electrons given by $\cos[\chi(\vec{k})]$
the measured photoelectrons have a polarization 
proportional to $\sin^{2}[\chi(\vec{k})]$ for photons given by the polarization $\vec{e}_{s=2}(\theta=0, \phi=0)$  and  $\cos^{2}[\chi(\vec{k})]$ for $\vec{e}_{s=2}(\theta, \phi=0)$.
The two polarizations are shown in figures $2$ and $3$.

To conclude  a model  for   computing  the photoelectrons intensity  and polarization   has been introduced.  We  show that the polarization of the photoelectrons depends on the laser polarization.
\begin{figure}
\begin{center}
\includegraphics[width=6.0 in ]{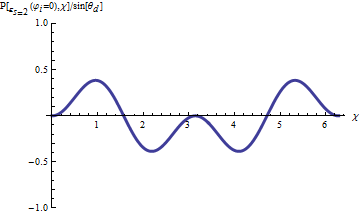}
\end{center}
\caption{The photoelectrons polarization $ P^{(2)}[\chi(\vec{k})=\phi_{d},\vec{e}_{s=2}(\theta=0, \phi=0),\theta_{d},\phi_{d}]$ for the  photon  polarization   $ \vec{e}_{s=2}(\theta=0, \phi=0)$ as a function the surface electron polar angle $ \chi(\vec{k})=\phi_{d}$ .} 
\end{figure} 
\clearpage
\begin{figure}
\begin{center}
\includegraphics[width=6.0 in ]{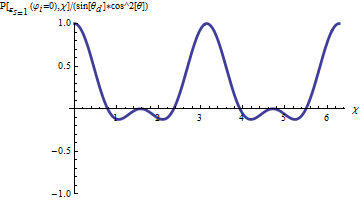}
\end{center}
\caption{The  photoelectrons  polarization $ P^{(2)}[\chi(\vec{k})=\phi_{d},\vec{e}_{s=1}(\theta, \phi=0),\theta_{d},\phi_{d}]$  for the polarization photon  $ \vec{e}_{s=1}(\theta, \phi=0)$ as a function the surface electron polar angle $ \chi(\vec{k})=\phi_{d}$ } 
\end{figure}

\end{document}